\documentclass[a4paper,11pt]{article}
\usepackage[utf8]{inputenc}
\usepackage[english]{babel}
\usepackage{braket}
\usepackage{xspace}
\usepackage{bbm}
\usepackage{mathdots}
\usepackage{stackrel}
\usepackage{xcolor}
\usepackage{mathtools}
\usepackage{bbm}
\usepackage{subfigure}
\usepackage[T1]{fontenc}               
\usepackage[left=3cm,
            right=2.5cm,
            top=2.5cm,
            bottom=3cm
            ]{geometry}                
\usepackage{graphicx}
\usepackage{mathrsfs}
\usepackage{appendix}
\usepackage{fancyhdr}
\usepackage{amsmath,                   
            amssymb,                   
            amsthm}                    
\usepackage{cite}

\usepackage{setspace} 

\usepackage[colorlinks=true,linkcolor=blue, citecolor=blue, bookmarks]{hyperref}

\numberwithin{equation}{section}

\def \be {\begin{equation}} 
\def \ee {\end{equation}} 
\def \l {\left(} 
\def \r {\right)} 
\def \la {\langle} 
\def \ra {\rangle}  

\date{}

\title{R\'enyi entropy and negativity for massless complex boson at conformal interfaces and junctions}

\author{Luca Capizzi$^1$, Sara Murciano$^1$, and Pasquale Calabrese$^{1,2}$}
\begin{document}
\maketitle

$^1$SISSA and INFN Sezione di Trieste, via Bonomea 265, 34136 Trieste, Italy.\\
$^{2}$International Centre for Theoretical Physics (ICTP), Strada Costiera 11, 34151 Trieste, Italy.\\
\begin{abstract} 
We consider the ground state of a theory composed by $M$ species of massless complex bosons in one dimension coupled together via a conformal interface. We compute both the R\'enyi entropy and the negativity of a generic partition of wires, generalizing the approach developed in a recent work for free fermions. These entanglement measures show a logarithmic growth with the system size, and the universal prefactor depends both on the details of the interface and the bipartition. We test our analytical predictions against exact numerical results for the harmonic chain.

\end{abstract}

 \tableofcontents

\section{Introduction}
The characterisation of the entanglement of many-body quantum systems is an intense research theme nowadays \cite{afov-08,intro2}. This interest emerges from different communities, ranging from quantum information \cite{eisert-2010,Laflorencie-16} to quantum gravity \cite{bkls-86,Srednicki-93,rt-06,Raamsdonk,maldacena,hlw-94,cw-94,ct-16,ot-15}, to get to condensed matter: it is now understood
that extended quantum systems and their phases may be characterised through their
entanglement properties. Given its relevance, entanglement is often used as a probe to investigate the features of a system, as we will do also in this manuscript. Before doing that, it is worth defining a way to quantify it: starting from a pure state $|\Phi\rangle$ and a bipartition $A \cup B$, the subsystem $A$ is described by the reduced density matrix 
$\rho_A=\mathrm{Tr}_B |\Phi\rangle\langle \Phi|$ and the entanglement can be quantified by the R\'enyi entropies \cite{cc-04,cc-09}
\begin{equation}\label{eq:renyidef}
    S_n(A)=\frac{1}{1-n}\log \mathrm{Tr}\l\rho_A^n\r.
\end{equation}
For $n\to 1$ we recover the von Neumann entropy $S(A) = - \text{Tr}\l \rho_A \log \rho_A \r$, often called just {\it entanglement entropy}. In a previous paper \cite{cmc-22}, we studied the scaling of the R\'enyi entropies and negativities of the ground state of a (1+1)-dimensional conformal field theory (CFT) built with $M$ species of massless free Dirac fermions coupled at one boundary point via a 
conformal junction/interface. We were motivated by the interest towards this class of critical systems characterised by entangling points which are partially transmitting and reflecting \cite{coa-03,kf-92,p-051,coc-13,p-05,ep-10,cmv-12,ep-12,ep2-12,rpr-22}, whose behaviour is usually understood through the Boundary Conformal Field Theory (BCFT) \cite{al-91,cardy-84,bcft,c-04,s-98,kondo,p-96,bddo-02,bbr-13,kmcm-21}. In particular, each specie is represented by a wire of finite length $L$ and we have computed the R\'enyi entropies of a generic bipartition of wires, finding a logarithmic growth with $L$. Our results generalised the main finding of \cite{ss-08}, valid for two wires only, which is
\be
S(A) = \frac{c_{\text{eff}}(\sqrt{\mathcal{T}})}{6}\log \frac{L}{\varepsilon} + \dots,
\label{eq:VNentropySS}
\ee
where $\mathcal{T}$ represents the transmission probability and $\varepsilon$ an ultraviolet cutoff. The parameter $c_{\text{eff}}$, dubbed as  \textit{effective central charge}, is 
a monotonic function of its argument satisfying
\be
c_{\text{eff}}(0)=0, \quad c_{\text{eff}}(1) = c.
\ee
Here $c$ is the central charge, and explicit functional form of $c_{\text{eff}}(\sqrt{\mathcal{T}})$ depends both on the theory and on the details of the interface.
The study of the R\'enyi entropy for a Dirac field has allowed us to develop a systematic strategy that we will apply also in the present work for $M$ species of a free complex boson. 

Therefore, the purpose of this manuscript is two-fold. Firstly, we show how one can treat the bosonic case in complete analogy with the fermionic one. Thus, we consider a junction of $M$ wires, one per species, coupled by a conformal invariant scattering matrix $S$ (see \cite{coa-06,bm-06,bms-07,hc-08,bcm-09,bms-09,cmr-13,cmv-11,cmv-12a}), and we study the entanglement of a generic bipartition among the wires of the junctions. On the field theory side, we apply the folding trick which turns the problem of constructing conformal interfaces into one of constructing boundary states. This leads to the computation of charged partition functions where the implementation of the $U(1)$ symmetry for a complex scalar field is different from the topological charge of the complex fermion after bosonisation: this technical but relevant distinction is the origin of the different results we present in the manuscript. Then, we also take into account the lattice version of the problem, which is a system of coupled chains of oscillators (aka harmonic chain), checking the results derived in the continuum limit.
Second, the presence of several wires (or bosonic species) is an ideal scenario to study the entanglement between non complementary parts and an useful entanglement measure in this context is the {\it negativity} \cite{vidal,plenio-2005}. Its computation requires the implementation of the partial transposition of the reduced density matrix: this operation turns out to be rather involved for fermionic systems, since the partial transpose of Gaussian fermionic density matrices is not Gaussian \cite{ssr-17,srrc-19,ez-15,srgr-18,sr-19,sr-19a,mbc-21,csg-19,mvdc-22} and, therefore, 
in \cite{cmc-22} we have adopted a definition which is more suitable for free fermions. Here, since we deal with bosons, we can apply the standard notion of partial transposition, that we report for completeness. Let us fix a tripartition of the $M$ wires into $A \cup B \cup C$, containing $M_A, M_B, M_C$ wires, respectively. Given the reduced density matrix $\rho_{A \cup B}= \mathrm{Tr}_C \l\rho_{A \cup B \cup C}\r$, we can express its matrix elements  as
\begin{equation}
\rho_{A\cup B}=\sum_{ijkl}\braket{e^A_i,e^B_j|\rho_{A\cup B}|e^A_k, e^B_l}\ket{e_i^A,e_j^B}\bra{e^A_k,e^B_l},
\end{equation}
where $\ket{e_j^A}$ and $\ket{e_k^B}$ are orthonormal bases in the Hilbert spaces $\mathcal{H}_A$ and $\mathcal{H}_B$ corresponding to the $A$ and $B$ regions, respectively. The partial transpose for the subsystem $B$ ($T_B$) is defined by exchanging the matrix elements in the subsystem $B$, i.e.
\begin{equation}
\label{eq:bosonic}
\rho^{T_B}_{A\cup B}=\sum_{ijkl}\braket{e^A_i,e^B_l|\rho_{A\cup B}|e^A_k, e^B_j}\ket{e_i^A,e_j^B}\bra{e^A_k,e^B_l}.
\end{equation} 
The entanglement between $A$ and $B$ is provided by the (logarithmic) negativity, defined as
\begin{equation}
    \mathcal{E}=\log ||\rho_{A \cup B}^{T_B}||,
\end{equation}
where $|| \cdot ||$ is the trace norm (that does not depend on the choice of the basis). This quantity can be also derived by properly implementing the replica limit: starting from the R\'enyi negativities
\begin{equation}
    \mathcal{E}_n=\log (\rho_{A\cup B}^{T_B})^n,
\end{equation}
we need to consider their associated analytic continuation for even powers of $n$ ($n=n_e$) and the negativity is given by the limit \cite{cct-12,cct-13}
\begin{equation}
    \mathcal{E}=\lim_{n_e \to 1} \mathcal{E}_{n_e}.
\end{equation}
We find that the R\'enyi negativities, as well as the logarithmic negativity, show a logarithmic growth in $L$ with a universal prefactor depending on the details of the interfaces and the bipartition (as already observed in \cite{cmc-22} for fermions).\\


Before embarking in the technical computations, we provide a bird's eye view about the structure of this manuscript: in Sec. \ref{sec:CFTapproach}, we review the construction of bosonic boundary states in CFT, which is the starting point to compute the R\'enyi entropies between complementary sets of wires. We extend this formalism to compute the negativity in Sec. \ref{sec:negativity}. Then, in Sec. \ref{sec:num} we benchmark our CFT predictions against the numerical data obtained for the complex harmonic chain. We leave our concluding remarks to Sec. \ref{sec:conlusions}.

\section{CFT approach: description of the method and the application to the R\'enyi entropy}\label{sec:CFTapproach}

We briefly summarize the CFT approach for the evaluation of the R\'enyi entropies in conformal junctions of bosonic CFTs, following mainly the setting of Ref. \cite{cmc-22}, which generalises systematically the strategy of \cite{bddo-02,ss-08,gm-17,bb-15}.
We consider $M$ wires of length $L$, each one described by a CFT that we label as $
\text{CFT}_j, \, j=1,\dots,M$ joined together at the point $x=0$. By applying the folding trick, we can work with a single CFT, denoted as
\be
\text{M-CFT} = \text{CFT}_1 \otimes \dots \otimes \text{CFT}_M
\ee
that in Euclidean space-time corresponds to a single infinite strip of width $L$. In terms of the space-time complex coordinate $w$, the joining between the different wires is specified by the boundary conditions along the lines
\be
{\rm Re}(w) = 0, \quad {\rm Re}(w) = L.
\ee
At ${\rm Re}(w) = L$, we choose boundary conditions that decouple the wires: as a consequence, they do not affect the correlation properties among distinct wires and we can avoid to specify the details about its structure. On the other hand, at the defect ${\rm Re}(w) = 0$, we assume that the boundary conditions couple explicitly different wires and we denote the corresponding boundary state by $\ket{S}$. In order to regularise the ultraviolet  and infrared divergences we cut the strip along $\varepsilon<|w|<L$, with $\varepsilon$ a cutoff, and then, though the conformal transformation
\be
z=\log w,
\ee
we map this portion of the strip
onto
\be
\mathrm{Re} (z) \in [\log \varepsilon,\log L]
 \quad \mathrm{Im}(z) \in [-\pi/2,\pi/2].
\ee
The partition function in this geometry can be written as
\be
\mathcal{Z} = \bra{S}\exp\l -\pi H \r\ket{S},
\label{ZvsS}
\ee
where the Hamiltonian is 
\be
H = \frac{2\pi}{\log \frac{L}{\varepsilon}}\l L_0 + \bar{L}_0 \r,
\ee
with $L_0,\bar{L}_0$ being generators of the Virasoro algebra of the $\text{M-CFT}$. 

In the remainder of the manuscript, we will characterise the boundary states of the free complex bosons and compute explicitly the partition function \eqref{ZvsS} in the presence of branch-cuts, which are directly related to the entanglement measures.

\subsection{Boundary states for free bosons}

In this section, we review the characterisation of the boundary states of a multispecies free complex boson, following \cite{bddo-02,gm-17}. We emphasise that a complex bosonic system can be expressed as two copies of a real bosonic one, thus one can generalise the results available for real bosons via a doubling of the degrees of freedom. In particular, a theory made by $M$ complex bosonic species has a total central charge
\be
c = 2M,
\ee
since each real bosonic specie carries a central charge $c=1$ \cite{DiFrancesco-97}. We denote the left/right components of the $j$-th species of the field $\Phi$ as
\be
\Phi^j(z), \quad \bar{\Phi}^j(\bar{z}), \quad j=1,\dots,M,
\ee
and, similarly, for the antiparticle field $\Phi^\dagger$ we use the following symbols
\be
(\Phi^\dagger)^j(z), \quad (\bar{\Phi}^\dagger)^j(\bar{z}), \quad j=1,\dots,M.
\ee
Expanding the fields in their Laurent modes via radial quantisation \cite{DiFrancesco-97}, it is possible to extract the creation/annihilation operator. In fact, to each mode, parametrised by a natural number $
k \in \mathbb{N},$
we can associate an operator $\Phi^j_{-k}$, which creates a left-moving particle of the $j$-th species, and an annihilation operator $\Phi^j_{k}$ that destroys it. For the sake of convenience, we normalise the Fourier modes so that their commutation relations are
\be
[\Phi^j_{k},\Phi^{j'}_{-k'}] = \delta_{j,j'}\delta_{k,k'}, \quad k,k'>0, \quad j,j'=1,\dots,M.
\ee
The same construction applies for the right modes, and similarly to the left/right modes of the antiparticle field $\Phi^\dagger$. The zero-modes of the fields commute among each other
\be
[\Phi^j_{0},\Phi^{j'}_{0}]=0, \quad j,j'=1,\dots,M
\ee
as explained in \cite{DiFrancesco-97}, and with the other modes, thus they can be regarded as constants of motion. For the sake of simplicity, we will assume that
\be
\Phi^j_0\ket{0} = \bar{\Phi}^j_0\ket{0}= (\Phi^\dagger)^j_0\ket{0} = (\bar{\Phi}^\dagger)^j_0\ket{0}= 0,
\ee
which amounts to neglect the presence of the zero modes. This technical assumption does not spoil the leading behaviour of the entanglement measures we are interested in, and simplifies the calculations (see \cite{gm-17,bddo-02} for further details about the zero modes contributions).\\
Let us consider the explicit construction of the boundary state mixing the junctions in terms of the bosonic modes. We focus on boundary conditions which preserve the global $U(1)$ symmetry
\be
\Phi^j \rightarrow e^{i\theta}\Phi^j, \quad  (\Phi^\dagger)^j \rightarrow e^{-i\theta}(\Phi^\dagger)^j,
\ee
corresponding to the imbalance of particles and antiparticles. In this case, a set of boundary conditions which keep the action of the bosonic theory quadratic is parametrised by a unitary matrix $S$ such that
\be
S \in U(M),
\ee
and the associated boundary states $\ket{S}$ satisfy
\be
(\Phi^i_{k} - S_{ij}\Phi^j_{-k})\ket{S} = (\Phi^i_{k} - \bar{S}_{ij}\Phi^j_{-k})\ket{S} =0.
\ee
Here $\bar{S}$ is the matrix complex conjugated to $S$ and the sum over $j$ is implicit. A solution to the previous relation is provided by
\be
\ket{S} = \prod_{k>0}\exp\l S_{jj'}{\Phi^\dagger}^j_{-k}\overline{\Phi}^{j'}_{-k} + (\Phi \leftrightarrow \Phi^\dagger)\r\ket{0}.
\label{eq_BosonBstate}
\ee

\subsection{R\'enyi entropies for a generic bipartition between wires}

We now consider a subset $A$ of $M_A\leq M$ wires and we describe how to compute the R\'enyi entropies using the replica trick. The key objects are the moments of the reduced density matrix $\rho_A$, expressed by \cite{cc-04,cc-09}
\be
\text{Tr}(\rho^n_A) = \frac{\mathcal{Z}_n}{\mathcal{Z}_1^n},
\ee
with $\mathcal{Z}_n$ being a partition of the $n$-replicated theory with branch-cuts along $A$ and $\mathcal{Z}_1$ the partition function of the single replica theory.\\
Before proceeding with the calculations, we briefly mention that, for the complex free bosonic theory, it is possible to relate $\mathcal{Z}_n$ to a $U(1)$-charged single-replica partition function, as explained below. Given the vacuum $\ket{0}$ of the theory and $Q_A$ the generator of the $U(1)$ symmetry
restricted to the subsystem $A$, we define the $U(1)$ charged partition function as
\be
\mathcal{Z}_1(\alpha) = \bra{0}e^{i\alpha Q_A}\ket{0}.
\ee
Using the property that the action is quadratic in the bosonic field, and that $\mathcal{Z}_n$, written as a Gaussian integral, eventually decouples as a product of $n$ Gaussian integrals, it was shown in \cite{ch-05} that it is possible to write the following factorisation
\be
\mathcal{Z}_n = \prod^{n-1}_{p = 0} \mathcal{Z}_1\l \alpha = \frac{2\pi p}{n} \r.
\label{eq:Zn_product}
\ee
We mention that a similar decomposition in terms of products over $U(1)$ partition functions 
appeared also in the computation of the symmetry resolved entanglement \cite{gs-18,xas-18,brc-19,mdc-20,ccdms-22a,ccdms-22b,amc-22}.
While Ref. \cite{ch-05} focuses on the ground state without boundaries, the same considerations can be applied here, being the boundary states in Eq. \eqref{eq_BosonBstate} Gaussian, too. Hence, coming back to our original problem, we consider the $\text{M-CFT}=\text{CFT}_1\otimes \text{CFT}_2 \dots \otimes \text{CFT}_M$, we replicate it $n$-times, ending up into $\text{M-CFT}^{\otimes n}$, and we evaluate the partition function $\mathcal{Z}_n$ as a product of $U(1)$ charged partition functions. In our specific case, we have
\be
\mathcal{Z}_1(\alpha) = \bra{S}e^{i\alpha Q_A}q^{L_0+\bar{L}_0}\ket{S},
\ee
where the modular parameter $q$ is
\be
q \equiv \exp\l -\frac{2\pi^2}{\log\l L/\varepsilon \r} \r.
\ee
Eventually, $\mathcal{Z}_n$ is given by Eq. \eqref{eq:Zn_product}, as in the absence of boundaries.\\
We now proceed with the computation of $\mathcal{Z}_1(\alpha)$ given by
\begin{multline}
\mathcal{Z}_1(\alpha) = \prod_{k >0}\bra{0} \exp\l  (S^\dagger)_{jj'}\overline{\Phi}^{j}_{k} {\Phi^\dagger}^{j'}_{k}+ (\Phi \leftrightarrow \Phi^\dagger)\r q^{L_0+\bar{L}_0}e^{i\alpha Q_A} \times \\
\exp\l S_{jj'}{\Phi^\dagger}^j_{-k}\overline{\Phi}^{j'}_{-k} + (\Phi \leftrightarrow \Phi^\dagger)\r\ket{0}.
\label{eq:U(1)Renyi}
\end{multline}
Here $L_0$ and $\bar{L}_0$ act on the creation operators of $\Phi$ (or equivalently $\Phi^\dagger$) as
\be
[L_0,\Phi^j_{-k}] = k\Phi^j_{-k}, \quad [L_0,\bar{\Phi}^j_{-k}] = k\bar{\Phi}^j_{-k}.
\label{eq:Vir_Comm}
\ee
We can decompose the scattering matrix $S$ in a block form
\be
S = \begin{pmatrix}S_{AA} & S_{AB} \\ S_{BA} & S_{BB}\end{pmatrix},
\label{eq:S_block}
\ee
with $A$ standing for the $M_A$ species belonging to the subsystem $A$, while $B$ refers to the remaining $M_B = M-M_A$ species. Then, we can split the set of indices $j=1,\dots,M_A+M_B$, associated to all the species, in the following two sets
\be
a=1,\dots,M_A, \quad b=1,\dots,M_B
\ee
to shorthand the species of $A$ and $B$ respectively. This allows us to write concisely the $U(1)$ charge $Q_A$, representing the imbalance between  particles and antiparticles in $A$, as
\be
Q_{A} = \sum_{k>0} \Phi^a_{-k}\Phi^a_{k} + \bar{\Phi}^a_{-k}\bar{\Phi}^a_{k} - (\Phi\rightarrow \Phi^\dagger),
\label{eq:Charge_Comm}
\ee
where the sum over $a$ is implicit. As a consequence of the commutation relations \eqref{eq:Vir_Comm} and of Eq. \eqref{eq:Charge_Comm}, it is possible to derive straightforwardly
\be
\begin{split}
& q^{L_0+\bar{L}_0}e^{i\alpha Q_A} {\Phi^\dagger}^a_{-k} = e^{-i\alpha } q^{k}{\Phi^\dagger}^a_{-k}q^{L_0+\bar{L}_0}e^{i\alpha Q_A},\\
& q^{L_0+\bar{L}_0}e^{i\alpha Q_A} {\Phi^\dagger}^b_{-k} =  q^{k}{\Phi^\dagger}^b_{-k}q^{L_0+\bar{L}_0}e^{i\alpha Q_A},\\
& q^{L_0+\bar{L}_0}e^{i\alpha Q_A} \bar{\Phi}^a_{-k} =  e^{i\alpha}q^{k}\bar{\Phi}^a_{-k} q^{L_0+\bar{L}_0}e^{i\alpha Q_A},\\
& q^{L_0+\bar{L}_0}e^{i\alpha Q_A} \bar{\Phi}^b_{-k} =  q^{k}\bar{\Phi}^b_{-k} q^{L_0+\bar{L}_0}e^{i\alpha Q_A}.
\end{split}
\label{eq:CommRel}
\ee
We now focus on the contribution of a single bosonic mode $k$ of field $\Phi$, namely we evaluate
\begin{multline}
\bra{0} \exp\l  (S^\dagger)_{jj'}\overline{\Phi}^{j}_{k} {\Phi^\dagger}^{j'}_{k}\r q^{L_0+\bar{L}_0}e^{i\alpha Q_A} \exp\l  S_{jj'}{\Phi^\dagger}^j_{-k}\overline{\Phi}^{j'}_{-k} \r\ket{0} = \\
\bra{0} \exp\l  (S^\dagger)_{jj'}\overline{\Phi}^{j}_{k} {\Phi^\dagger}^{j'}_{k}\r \times\\
\exp\l  q^{2k}S_{aa'}{\Phi^\dagger}^a_{-k}\overline{\Phi}^{a'}_{-k} +  q^{2k}S_{bb'}{\Phi^\dagger}^b_{-k}\overline{\Phi}^{b'}_{-k} +  e^{-i\alpha}q^{2k}S_{ab'}{\Phi^\dagger}^a_{-k}\overline{\Phi}^{b'}_{-k} +  e^{i\alpha}q^{2k}S_{ba'}{\Phi^\dagger}^b_{-k}\overline{\Phi}^{a'}_{-k} \r\ket{0},
\end{multline}
where the property $q^{L_0+\bar{L}_0}e^{i\alpha Q_A}\ket{0} = \ket{0}$ has been used. This expression can be computed as (see Eq. \eqref{eq_BoundOverlap} in the Appendix)
\begin{multline}
\bra{0} \exp\l  (S^\dagger)_{jj'}\overline{\Phi}^{j}_{k} {\Phi^\dagger}^{j'}_{k}\r q^{L_0+\bar{L}_0}e^{i\alpha Q_A} \exp\l S_{jj'}{\Phi^\dagger}^j_{-k}\overline{\Phi}^{j'}_{-k} \r\ket{0} = \\
\text{det}^{-1}\l \begin{pmatrix} 1 & 0 \\ 0 & 1\end{pmatrix}- q^{2k}\begin{pmatrix} S^\dagger_{AA} & S^\dagger_{BA} \\ S^\dagger_{AB}  & S^\dagger_{BB}  \end{pmatrix} \begin{pmatrix} S_{AA} & e^{-i\alpha}S_{AB} \\ e^{i\alpha}S_{BA}  & S_{BB}  \end{pmatrix}\r.
\label{eq:Block_det}
\end{multline}
Similarly, thanks to the unitarity of $S$, one can show (see \cite{cmc-22} for similar calculations)
\begin{multline}
\text{det}^{-1}\l \begin{pmatrix} 1 & 0 \\ 0 & 1\end{pmatrix}- q^{2k}\begin{pmatrix} S^\dagger_{AA} & S^\dagger_{BA} \\ S^\dagger_{AB}  & S^\dagger_{BB}  \end{pmatrix} \begin{pmatrix} S_{AA} & e^{-i\alpha}S_{AB} \\ e^{i\alpha}S_{BA}  & S_{BB}  \end{pmatrix}\r\propto \\
 \text{det}^{-1}\l 1 - 2(S_{AA}^\dagger S_{AA} + (1-S_{AA}^\dagger S_{AA})\cos \alpha )q^{2k} +q^{4k} \r,
\end{multline}
up to an irrelavant $\alpha$-independent proportionality constant. Combining the contributions coming from all the modes $k$ of $\Phi$ and $\Phi^\dagger$, one arrives at the following closed expression for the charged partition function
\be
\mathcal{Z}_1(\alpha) \propto \prod_{k>0} \text{det}^{-2}\l 1 - 2(S_{AA}^\dagger S_{AA} + (1-S_{AA}^\dagger S_{AA})\cos \alpha )q^{2k} +q^{4k} \r.
\label{eq:U(1)1replica}
\ee
Summing over the $n$ values of the $U(1)$ flux, we can finally write the $n$-sheeted partition function $\mathcal{Z}_n$ in Eq. \eqref{eq:Zn_product} as
\begin{multline}
\mathcal{Z}_n = \prod^{n-1}_{p=0}\mathcal{Z}_1(\alpha = 2\pi p/n) \propto \\ 
\prod^{n-1}_{p=0}\prod_{k >0} \text{det}^{-2}\l 1-2(S_{AA}^\dagger S_{AA} + (1-S_{AA}^\dagger S_{AA})\cos (2\pi p/n))q^{2k}+q^{4k}\r.
\label{eq:Zn_PartFun}
\end{multline}
From the result above,
one easily realises that, whenever $M_A \geq 1$, $\mathcal{Z}_n$ can be written as a product of $M_A$ factorised contributions corresponding to the eigenvalues of the $M_A\times M_A$ matrix $1-S^\dagger_{AA} S_{AA}$. Indeed, for each eigenvalue $\mathcal{T}_a \in \text{Spec}(1-S^\dagger_{AA}S_{AA})$ we introduce the following quantity
\be
\mathcal{Z}_{n,\mathcal{T}_a} \equiv \prod^{n-1}_{p=0}\prod_{k >0} \text{det}^{-2}\l 1-2((1-\mathcal{T}_a) + \mathcal{T}_a\cos (2\pi p/n))q^{2k}+q^{4k}\r,
\label{eq:Zn_T}
\ee
so that $\mathcal{Z}_{n}$ can be written as
\be
\mathcal{Z}_{n} = \prod_{a=1}^{M_A} \mathcal{Z}_{n,\mathcal{T}_a}.
\ee
Plugging this relation into the definition of the R\'enyi entropy of Eq. \eqref{eq:renyidef}, we obtain
\be\label{eq:rcft}
S_{n}(A) = \sum^{M_A}_{a=1} S_{n,\mathcal{T}_a},
\quad 
S_{n,\mathcal{T}_a} = \frac{1}{1-n}\log \frac{\mathcal{Z}_{n,\mathcal{T}_a}}{\mathcal{Z}^n_{1,\mathcal{T}_a}},
\ee
where $S_{n,\mathcal{T}_a}$ is the contribution coming from the eigenvalue $\mathcal{T}_a$, which can be regarded as an effective transmission probability.\\
We proceed further in order to provide a closed expression for the leading term of the entropies in the relevant limit $q\rightarrow 1$, corresponding to $\frac{L}{\varepsilon} \rightarrow \infty$. For future convenience, we define a parameter $\alpha'$, depending on $\alpha$ and $\mathcal{T}_a$, as follows
\be
2\cos \alpha' \equiv 2(1-{\cal T}_a + {\cal T}_a\cos \alpha ),
\label{eq:alphaprime}
\ee
a relation which appeared already in Ref. \cite{cmc-22}. Comparing the definition of $\mathcal{Z}_{1,\mathcal{T}_a}$ in Eq. \eqref{eq:Zn_T} with the properties of the Jacobi Theta functions (discussed in the Appendix \eqref{sec_Jacobi}), one can write
\be
\mathcal{Z}_{1,{\cal T}_a}(\alpha) \propto \l \frac{\sin(\alpha'/2)}{\theta_1\l \frac{\alpha'}{2\pi},q \r}\r^2,
\ee
up to an irrelevant $\alpha$-independent constant. In the limit $q\rightarrow 1$, the previous expression can be further simplified yielding
\be
\log \frac{\mathcal{Z}_{1,{\cal T}_a}(\alpha)}{\mathcal{Z}_{1,\mathcal{T}_a}(0)} \simeq -\l\frac{\alpha'}{2\pi} - \l\frac{\alpha'}{2\pi}\r^2\r \log \frac{L}{\varepsilon}, \quad \alpha \in [0,2\pi],
\ee
which is the main result of this section. We stress that the dependence on $\alpha'$ is distinct from the one found in \cite{cmc-22}, as a consequence of the different implementation of the $U(1)$ symmetry in the complex scalar theory.
For each eigenvalue $\mathcal{T}_a$, the relation between $\alpha'$ and $\alpha$, defined by \eqref{eq:alphaprime}, can be inverted as
\be
\alpha' = 2 \text{arcsin}\l\sqrt{\mathcal{T}_a} \sin \frac{\alpha}{2}\r.
\ee
and the contribution to the total entropy in Eq. \eqref{eq:rcft} from each transmission probability reads
\be\label{eq:singlecontr}
S_{n,\mathcal{T}_a} =\log\frac{L}{\varepsilon}\frac{1}{n-1}\sum^{n-1}_{p=0} \left[ \frac{1}{\pi}\text{arcsin}\l\sqrt{\mathcal{T}_a} \sin \frac{\pi p}{n} \r - \l\frac{1}{\pi}\text{arcsin}\l\sqrt{\mathcal{T}_a} \sin \frac{\pi p}{n} \r\r^2\right].
\ee
As a sanity check, we specialise this prediction to the case $M=2$, $M_A=1$, and
\be
S = \begin{pmatrix} 0 & 1 \\ 1 & 0\end{pmatrix},
\ee
which describes a completely transmissive interface between two wires. Here $S_{AA} =0$ is just a number and the only eigenvalue of $1-S^\dagger_{AA}S_{AA}$ is thus
\be
\mathcal{T}_a = 1.
\ee
Inserting this value of $\mathcal{T}_a$ in \eqref{eq:alphaprime} one finds $\alpha' = \alpha$. Putting all the pieces together we get
\be
\begin{split}
S_{n}(A) = &\frac{1}{1-n}\sum^{n-1}_{p=0}\log \frac{Z_{1,\mathcal{T}_a}(\alpha=2\pi p/n)}{Z_{1,\mathcal{T}_a}(0)} \simeq \log \frac{L}{\varepsilon}\frac{1}{n-1}\sum^{n-1}_{p=0} \l \frac{p}{n}\r - \l \frac{p}{n}\r^2 =\\
& \frac{1}{6}\l 1+ \frac{1}{n}\r \log \frac{L}{\varepsilon}.
\end{split}
\label{eq:2wire_T1}
\ee
For $n\rightarrow 1$, this is compatible with the general prediction available for a boundary CFT of central charge $c=2$ (i.e. the complex boson)
\be
S_1(A) \simeq \frac{1}{3}\log \frac{L}{\varepsilon}.
\ee
Similarly, for a partially trasmissive interface with scattering matrix
\be
S = \begin{pmatrix} -\sqrt{1-\mathcal{T}} & \sqrt{\mathcal{T}}\\ \sqrt{\mathcal{T}} & \sqrt{1-\mathcal{T}} \end{pmatrix},
\ee
the only eigenvalue of $1-S^\dagger_{AA}S_{AA}$ is precisely
\be
\mathcal{T}_a = \mathcal{T},
\ee
with $\mathcal{T}$ the parameter entering in the definition of the scattering matrix $S$. 
The R\'enyi entropy takes the following form
\be
S_{n}(A) \simeq \log\frac{L}{\varepsilon}\frac{1}{n-1}\sum^{n-1}_{p=0} \left[ \frac{1}{\pi}\text{arcsin}\l\sqrt{\mathcal{T}} \sin \frac{\pi p}{n} \r - \l\frac{1}{\pi}\text{arcsin}\l\sqrt{\mathcal{T}} \sin \frac{\pi p}{n} \r\r^2\right],
\label{eq:2Wire_T}
\ee
which gives back \eqref{eq:2wire_T1} once specialised to $\mathcal{T}=1$.\\
We emphasise that the result \eqref{eq:2Wire_T} already appeared in the literature \cite{ss-08} for the real boson (i.e. the prefactor of the logarithmic term is half of the one appearing in Eq. \eqref{eq:2Wire_T}), together with its analytical continuation for $n\rightarrow 1$ which is not reported here. Moreover, we remark that we focused on the leading term only, the one proportional to $\log \frac{L}{\varepsilon}$, while the $O(1)$ terms, which are important for the characterisation of the topological entropy (see \cite{ffrs-04,ffrs-07,aamf-16,bbjs-16,Jaud-2017,rs-22}), have been neglected. As a matter of fact, these terms are sensible to details of the vacuum appearing in the definition of \eqref{eq_BosonBstate}, and whether the boson is compact or noncompact. However, these subleading effects go beyond the scope of this work, and we do not discuss them further, referring the reader to \cite{ss-08} for additional details.

\section{Negativity}\label{sec:negativity}
As done in \cite{cmc-22}, here we exploit the presence of several wires in order to compute the entanglement between non complementary parts. For this purpose, we start from the R\'enyi negativity defined in terms of the partial transpose reduced density matrix reported in Eq. \eqref{eq:bosonic}. Then, we apply the replica limit to obtain the logarithmic negativity.

\subsection{R\'enyi negativity}
Let us start from the replica approach for the negativity computing the moments of the partial transpose reduced density matrix in terms of a ratio of partition functions as \cite{cct-12,cct-13,rac-16b}
\begin{equation}
    \mathrm{Tr}(\rho_{A \cup B}^{T_B})^n=\frac{\hat{ \mathcal{Z}}_n}{\hat{\mathcal{Z}}_1^n},
\end{equation}
where $\hat{Z}_n$ is the partition function
in the $n$-sheeted Riemann surface built by implementing the partial transpose in the subsystem $B$ in the path-integral formalism. These moments, which are the key ingredients here, are directly related to the negativity, and we mention that they appear also in the computation of other entanglement measures \cite{ekh-20,ncv-21}.\\
As already done for the R\'enyi entropies, the diagonalisation in the replica space allows us to write this ratio as a product of $n$ single-replica $U(1)$ charged partition functions. Here the twisting phase is $e^{i\alpha}$ for the subsystem $A$ and $e^{-i\alpha}$ in $B$. One can thus express the $n$-th R\'enyi negativity as
\be
\mathcal{E}_{n} 
= \sum^{n-1}_{p=0} \log \frac{\hat{\mathcal{Z}}_1(\alpha=2\pi p/n)}{\hat{\mathcal{Z}}_1(0)},
\label{eq:Renyi_Neg_Sum}
\ee
and the $U(1)$ partition function $\hat{\mathcal{Z}}_1(\alpha)$ entering in the definition of \eqref{eq:Renyi_Neg_Sum} is
\begin{multline}
\hat{\mathcal{Z}}_1(\alpha) \equiv \bra{S}q^{L_0+\bar{L}_0}e^{i\alpha Q_A -i\alpha Q_B}\ket{S} = \\
\prod_{k >0} \bra{0} \exp\l  (S^\dagger)_{jj'}\overline{\Phi}^{j}_{k} {\Phi^\dagger}^{j'}_{k} + (\Phi \leftrightarrow \Phi^\dagger)\r q^{L_0+\bar{L}_0}\times\\
e^{i\alpha Q_A-i\alpha Q_B} \exp\l S_{jj'}{\Phi^\dagger}^j_{-k}\overline{\Phi}^{j'}_{-k} + (\Phi \leftrightarrow \Phi^\dagger) \r\ket{0}.
\label{eq:U(1)negativity}
\end{multline}
The building block entering in our formula is thus
\begin{multline}\label{eq:det}
\bra{0} \exp\l  S_{jj'}\overline{\Phi}^{j}_{k} {\Phi^\dagger}^{j'}_{k}\r q^{L_0+\bar{L}_0}e^{i\alpha Q_A-i\alpha Q_B} \exp\l S_{jj'}{\Phi^\dagger}^j_{-k}\overline{\Phi}^{j'}_{-k} \r\ket{0} =\\
\text{det}^{-1}\l  1-  q^{2k}S \begin{pmatrix} S_{AA} & e^{-i2\alpha}S_{AB} & e^{-i\alpha}S_{AC} \\ e^{i2\alpha}S_{BA}  & S_{BB} & e^{i\alpha}S_{BC}\\
e^{i\alpha}S_{CA} & e^{-i\alpha}S_{CB} & S_{CC}
\end{pmatrix}\r,
\end{multline}
a relation which can be shown through the formula \eqref{eq_BoundOverlap}, as it has been done for \eqref{eq:Block_det}.\\
So far we did not specify explicitly the number of wires belonging to the bipartitions, and \eqref{eq:U(1)negativity} holds generically. To provide some compact analytical results, from now on we restrict to the following specific situations
\begin{itemize}
\item $M_A =M_B=M_C=1$, so that the total number of wires is $M=3$ and we compute the negativity between two of them;
\item $S$ is not only unitary but also Hermitian, which implies $S^2=1$ and its eigenvalues can be just $\pm 1$. This is not a very restrictive condition since the hermiticity is a necessary condition for some physical systems (see  for example\cite{bms-07}).
\end{itemize}
After some algebraic manipulations, using the unitarity of $S$ and taking the product over the Fourier modes, one can show that (see \cite{cmc-22} for similar computations)
\be\label{eq:tocheck}
\hat{\mathcal{Z}}_1(\alpha) \propto \prod_{k>0} (1-2\cos \hat{\alpha}'q^{2k}+q^{4k})^{-2},
\ee
up to an irrelevant $\alpha$-independent prefactor, with $\hat{\alpha}'$ being a function of the $S$-matrix and the flux $\alpha$ defined by
\be
2 \cos \hat{\alpha}' =   -1+ S^2_{AA} + S^2_{BB} + S_{CC}^2+(1+S^2_{CC} -S^2_{BB}-S^2_{AA})\cos(2\alpha) + 2(1-S^2_{CC})\cos \alpha.
\label{eq:alpha_tilde_pr}
\ee
We mention that Eq. \eqref{eq:alpha_tilde_pr} differs explicitly from the result of free fermions in Ref. \cite{cmc-22}, since an additional $U(1)$ phase factor in $\hat{\mathcal{Z}}_1(\alpha) $ in Eq. \eqref{eq:U(1)negativity} (related to a different definition of the negativity) was present there.\\ 
The same formal structure of $\mathcal{Z}_1(\alpha)$ appearing in \eqref{eq:U(1)1replica}, is found for $\hat{\mathcal{Z}}_1(\alpha)$ (Eq. \eqref{eq:tocheck}) up to the replacement $\alpha'\rightarrow \hat{\alpha}'$, and we get
\be
\log \frac{\hat{\mathcal{Z}}_{1}(\alpha)}{\hat{\mathcal{Z}}_{1}(0)} \simeq -\l\frac{\hat{\alpha}'}{2\pi} - \l\frac{\hat{\alpha}'}{2\pi}\r^2\r \log \frac{L}{\varepsilon}, \quad \alpha \in [0,2\pi],
\ee
which already provides a prediction of the R\'enyi negativity $\mathcal{E}_{n}$ 
\be
\mathcal{E}_{n_e} = \sum^{n-1}_{p=0} \log \frac{\hat{\mathcal{Z}}_{1}(2\pi p /n)}{\hat{\mathcal{Z}}_{1}(0)},
\label{eq:Renyi_neg_sum}
\ee
for integer values of $n$. By plugging the results \eqref{eq:alpha_tilde_pr} for $\hat{\alpha}'$ into 
Eq. \eqref{eq:Renyi_neg_sum}, we finally obtain
\begin{multline}\label{eq:cal-final}
{\cal E}_{n}=
\sum_{p=0}^{n-1} \Big( \frac{1}{4\pi^2}
 \arccos^2 \left(-1+S^2_{BB}+S^2_{AA}+
 (1+S^2_{CC} -S^2_{BB}-S^2_{AA})\cos(2\pi p/n)^2 \right. \\ \left. + (1-S_{CC}^2)\cos(2\pi p/n) \right) - \frac{1}{2\pi} \arccos \left(-1+S^2_{BB}+S^2_{AA}+\right. \\ \left. 
 (1+S^2_{CC} -S^2_{BB}-S^2_{AA})\cos(2\pi p/n)^2 + (1-S_{CC}^2)\cos(2\pi p/n) \right) \Big)  \log \frac{L}\varepsilon\,
\end{multline}
which is the main result of this subsection.\\
To conclude this part, we give some checks for the charged partition function $\hat{\mathcal{Z}}_1(\alpha)$ in specific simple cases. For instance, let us assume that the third wire is decoupled from $A$ and $B$ and then
\be
S^2_{CC} =1.
\ee
In this case $A$ and $B$ are coupled via a transmission probability
\be
1-S^2_{AA} = 1-S^2_{BB},
\ee
and $\hat{\alpha}'$ is given by
\be
2\cos \hat{\alpha}'= 2S^2_{AA}+2(1-S_{AA}^2)\cos(2\alpha).
\ee
One recognises that this is the same value one would obtain for the parameter $\alpha'$ in Eq. \eqref{eq:alphaprime}, once a flux $e^{i2\alpha}$ is inserted along $A$ in the $U(1)$ partition function $\mathcal{Z}_1(\alpha)$ involving $A$ and $B$ only. Indeed, in this specific limit the theory becomes invariant under the symmetry generated by $Q_A+Q_B$, being $C$ decoupled, and since
\be
e^{i\alpha Q_A -i\alpha Q_B} = e^{i2\alpha Q_A -i\alpha (Q_A+Q_B)},
\ee
the equivalence with the insertion of $e^{i2\alpha Q_A}$ is manifest.\\
We now consider $\alpha = \pi$, whose value is related to the $2$-nd R\'enyi negativity, and from \eqref{eq:Renyi_neg_sum} one gets
\be
\mathcal{E}_2 = \log \hat{\mathcal{Z}}_1(\pi).
\ee
In this case
\be
2\cos \hat{\alpha}' = 2S^2_{CC} -2(1-S^2_{CC}),
\ee
which is the same value we would get for a charged partition function with the insertion of $e^{i\pi Q_C}$ only. This is expected since it follows from the more general identity
\be
\text{Tr}((\rho^{T_B}_{AB})^{2}) = \text{Tr}((\rho_{AB})^{2}) = \text{Tr}((\rho_C)^{2}),
\ee
which relates the 2-R\'enyi negativity to the 2-R\'enyi entropy.

\subsection{Analytic continuation}

In order to evaluate the logarithmic negativity by applying the replica limit $n_e \to 1$, we first  provide an analytic continuation in $n$ of Eq. \eqref{eq:cal-final}. 
To this goal, the strategy we device is the same used in \cite{cmc-22}, up to slight modifications:
\begin{itemize}
    \item Using the identities of Appendix \ref{sec_Jacobi},
    the $U(1)$ partition function $\hat{\mathcal{Z}}_1(\alpha)$ in Eq. \eqref{eq:tocheck} can be expressed through an integral representation in the limit $q \to 1$ 
    \begin{multline}
    \log \frac{\hat{\mathcal{Z}}_1(\alpha)}{\hat{\mathcal{Z}}_1(0)}=-\sum_{k >0 }2\log [ (1-q^{2k})^{-2}(1-2\cos \alpha'q^{2k}+q^{4k})]\\ \simeq
    \frac{1}{\log q}\int_0^{\infty}\frac{dt}{t} [\log (1+2\cos \alpha't+t^{2})-2\log(1-t)];
    \end{multline}
    \item The sum over the values of fluxes \eqref{eq:cal-final} can now be performed inside the integral, using the trigonometric identities of \ref{sec:trig}, which leads to our analytic continuation.
\end{itemize}
In particular, we can write
\be
(1-2\cos \hat{\alpha}' t + t^{2})=\left(\cos(\alpha)+\frac{b-\sqrt{b^2-4 ac}}{2c}\right)\left(c\cos(\alpha)+\frac{b+\sqrt{b^2-4 ac}}{2}\right),
\ee
with $a,b,c$ being the following functions of the $S$ matrix and $t$
\be
a=1+2 (1-S_{AA}^2-S_{BB}^2)t+t^{2},\qquad b=2(S_{CC}^2-1)t,\qquad c=2(S^2_{AA} + S^2_{BB} - S_{CC}^2-1)t.
\ee
Using the integral representation for the product over the $k$ modes in the limit $q \to 1$ reported in \ref{sec_Jacobi}, we get
\begin{multline}\label{eq:negativity}
\mathcal{E}_{n} = \sum^{n-1}_{p=0}\log \frac{\hat{\mathcal{Z}}_1(\alpha)}{\hat{\mathcal{Z}}_1(0)}=- \frac{\log (L/\varepsilon)}{2\pi^2}\int_0^1 \frac{dt}{t}\\ \times \sum^{n-1}_{p=0}\left[ \log \left(\cos(2\pi p/n)+\frac{b-\sqrt{b^2-4 ac}}{2c}\right)\left(c\cos(2\pi p/n)+\frac{b+\sqrt{b^2-4 ac}}{2}\right)-2\log(1-t)\right]\\
=-\frac{\log (L/\varepsilon)}{\pi^2}\int_0^1 \frac{dt}{t}\times\\  \log \frac{ ((x_1 - \sqrt{x_1^2-c^2})^{n/2}-(
   x_1 + \sqrt{x_1^2-c^2})^{n/2}) ((x_2 - \sqrt{x_2^2-1})^{n/2}-(
   x_2 + \sqrt{x_2^2-1})^{n/2})}{2^{n}(1-t)^{n}},
\end{multline}
where
\be
x_1=\frac{b+\sqrt{b^2-4ac}}{2}, \qquad x_2=\frac{b-\sqrt{b^2-4ac}}{2c}.
\ee
Eq. \eqref{eq:negativity} is the desired analytic continuation. 
At this point, the negativity is simply obtained by taking $n_e \to 1$:
\begin{multline}
{\cal E}= -\frac{\log (L/\varepsilon)}{\pi^2}\int_0^1 \frac{dt}{t}\times   \\
\log \frac{ ((x_1 - \sqrt{x_1^2-c^2})^{1/2} - (
   x_1 + \sqrt{x_1^2-c^2 })^{1/2}) ((x_2 - \sqrt{x_2^2-1  })^{1/2} - (
   x_2 + \sqrt{x_2^2-1})^{1/2})}{2(1-t)}.
\end{multline}

 Let us conclude this section by providing some useful cross-checks of our result. In the limit in which $S^2_{CC}=1$, the wire $C$ decouples and one recovers the result for two wires, which is given by the R\'enyi entropy with $n=1/2$ ($A$ and $B$ now form a pure state) \cite{ge-20}. 
   In this case, Eq. \eqref{eq:negativity} simplifies as
\begin{equation}
    \mathcal{E}=-\frac{\log (L/\varepsilon)}{\pi^2}\int_0^1\frac{dt}{t}\log \bigg(\frac{\sqrt{1+t^2+2t(1-2S^2_{AA})}-2\sqrt{t(1-S^2_{AA})}}{1-t}\bigg).
\end{equation}
Performing the change of variables $t=e^{-2x}$ and setting $s=\sqrt{1-S^2_{AA}}$, we can rewrite the previous integral as
\be
\mathcal{E} = -\log(L/\varepsilon)\frac{2}{\pi^2}\int^\infty_0 dx \log \frac{\sqrt{\sinh^2 x+s^2}-s}{\sinh x}.
\ee
Let us $\kappa(s) = \mathcal{E}/\log(L/\varepsilon)$. Taking the derivative with respect to $s$, we get
\be
\kappa'(s) = \frac{2}{\pi^2}\int^{\infty}_0 dx \frac{1}{\sqrt{s^2+\sinh^2 x}},
\ee
which can be rewritten as
\be
\kappa'(s) = \frac{2}{\pi^2}\int^{\infty}_0 dy \frac{1}{\sqrt{1+y^2}\sqrt{s^2+y^2}},
\ee
after the change of variable $y= \sinh x$. It is possible to recognise an integral expression of the elliptic integral $K$ \cite{gr-94}, which leads to 
\be
\kappa'(s) = \frac{2}{\pi^2 s}K\l 1-\frac{1}{s^2}\r =  \frac{2}{\pi^2 }K(1-s^2).
\ee
We can thus express $\kappa(s)$ as
\be
\kappa(s) = \frac{2}{\pi^2} \int^s_0 \frac{ds'}{s'}K\l 1-\frac{1}{{s'}^2}\r = \frac{G_{3,3}^{3,2}\left(
\begin{array}{c}
 \frac{1}{2},\frac{1}{2},1 \\
 0,0,0 \\
\end{array}\Big| \frac{1}{s^2}
\right)}{2 \pi ^3}= s\frac{G_{3,3}^{2,3}\left(
\begin{array}{c}
 \frac{1}{2},\frac{1}{2},\frac{1}{2} \\
 0,0,-\frac{1}{2} \\
\end{array}\Big| s^2
\right)}{2 \pi ^3},
\ee
with $G$ being the Meijer G-function \cite{gr-94}.

\section{Numerical methods for the complex harmonic chain}\label{sec:num}

In this section we describe a lattice realisation of a junction made of $M$ wires hosting bosonic degrees of freedom, employing standard techniques for Gaussian states \cite{ep-09}. These exact computations are the tested for the CFT predictions of the
previous sections.

\subsection{Complex harmonic chain and correlation functions}

Let us consider a one-dimensional complex massless boson on the line $[0, L]$ with Hamiltonian:
\be\label{eq:Ham}
H = \int^L_0 dx \  \Pi^{\dagger}(x)\Pi(x)+ \l\partial_x \Phi^{\dagger}(x)\r\l\partial_x \Phi(x)\r,
\ee
with $\Pi(x), \Pi^{\dagger}(x)$ being the conjugated momentum associated to the bosonic fields $\Phi(x), \Phi^{\dagger}(x)$ respectively. At each boundary point $x=0,L$, we impose either Neumann or Dirichlet conditions, i.e.
\be
\partial_x\Phi(x) = 0 \ (\text{Neumann}), \quad \Phi(x) = 0 \ (\text{Dirichlet}).
\ee
To fix the ideas, we choose Dirichlet boundary conditions (bc) at $x=L$.
Using
\begin{equation}
    \Pi=\frac{\Pi^{(1)}+i\Pi^{(2)}}{\sqrt{2}}, \quad \Phi=\frac{\Phi^{(1)}+i\Phi^{(2)}}{\sqrt{2}},
\end{equation}
the Hamiltonian \eqref{eq:Ham} can be rewritten as the sum of two Hamiltonians for the real bosons  $H=H^{(1)}+H^{(2)}$, where
\be\label{eq:Hamreal}
H^{(j)} = \frac{1}{2}\int^L_0 dx \  \Pi^{(j)}(x)\Pi^{(j)}(x)+ \l\partial_x \Phi^{(j)}(x)\r\l\partial_x \Phi^{(j)}(x)\r, \qquad j=1,2.
\ee
In turn, each Hamiltonian admits as a lattice discretisation  the harmonic chain with nearest neighbor interactions \cite{cct-13}. Indeed, by taking $L/a=N+1$, where $a$ is the lattice spacing (from now on, we set $a=1$), we can replace 
\begin{equation}
    \Pi^{(j)}(x)\to p^{(j)}(x), \qquad \Phi^{(j)}(x) \to q^{(j)}(x), \quad j=1,2
\end{equation}
obtaining 
\begin{equation}
    H^{(j)} \to \frac{1}{2}\sum_{n=0}^{N+1}(p^{(j)}(x))^2+\frac{1}{2}\sum_{n=0}^{N}(q^{(j)}(x+1)-q^{(j)}(x))^2, \quad j=1,2
\end{equation}
Also in this case, we can fix Dirichlet bc at $x=N+1$ and Dirichlet or Neumann bc at $x=0$. As a consequence, the single-particle energies ($\omega_k$) and eigenstates ($\phi_k(x)$) read (see \cite{ep-12,cct-13})
\begin{itemize}
\item Dirichlet-Dirichlet: $\phi^{DD}_{k}(x) = \sqrt{\frac{2}{N+1}}\sin (\frac{\pi k\, x}{N+1}), \quad \omega^{DD}_k =2 \sin\left(\frac{\pi k}{2N+2}\right)$,\\
\item Neumann-Dirichlet: $\phi^{ND}_{k}(x) = \sqrt{\frac{2}{N+1/2}}\cos (\frac{\pi (k-1/2)}{N+1/2}(x-1/2)), \quad \omega^{ND}_k  = 2\sin\left(\frac{\pi (k-1/2)}{2N+1}\right)$.\\
\end{itemize}
The entanglement entropy follows from the matrix $C=XP$ containing the position and momentum correlations in the subsystem \cite{ep-09}.
The expression of these two-point functions in the ground state of the system is ($j=1,2$)
\be
\begin{split}
X_{DD/ND}(x,x') \equiv& \la q^{(j)}(x) q^{(j)}(x')\ra =  \sum_{k=1}^N \frac{\phi^{DD/ND}_k(x)\phi^{DD/ND}_k(x')}{2\omega_k^{DD/ND}}, \\
\quad P_{DD/ND}(x,x') \equiv& \la p^{(j)}(x) p^{(j)}(x')\ra =  \sum_{k=1}^N \omega_k^{DD/ND}\frac{\phi^{DD/ND}_k(x)\phi^{DD/ND}_k(x')}{2}.
\end{split}
\ee
We now consider a system consisting of $M$ bosonic species, and we introduce an index $l=1,\dots,M$, labelling the species (or, equivalently the wires). At $x=0$, we impose a set of scale-invariant boundary conditions defined in terms of a scattering matrix $S$, Hermitian and unitary, which mixes all the species. We keep instead Dirichlet bc at $x=N+1$ for all the species, and we eventually obtain the following correlators ($j=1,2$)
\be
\begin{split}\label{eq:xil}
X_{ll'}(x,x') \equiv &\la q^{(j)}_l(x)q^{(j)}_{l'}(x')\ra = \l \frac{1+S}{2} \r_{ll'}X_{ND}(x,x') + \l \frac{1-S}{2} \r_{ll'}X_{DD}(x,x'),\\
P_{ll'}(x,x') \equiv &\la p^{(j)}_l(x)p^{(j)}_{l'}(x')\ra = \l \frac{1+S}{2} \r_{ll'}P_{ND}(x,x') + \l \frac{1-S}{2} \r_{ll'}P_{DD}(x,x'),
\end{split}
\ee
where $x,x'=1 \dots N$. Here the matrices $\l \frac{1\pm S}{2} \r$ enter as the projectors over the eigenspaces of $S$ with eigenvalues $\pm 1$, which correspond to boundary conditions of Neumann/Dirichlet type at $x=0$. We mention that a similar structure for the correlators appearing in Eq. \eqref{eq:xil} was already found in a junction of free fermions in Ref. \cite{cmc-22}, and we refer the reader to it for further details.

\subsection{The R\'enyi entropy between two arbitrary sets of bosonic species}
\begin{figure}[t]
\centering
	\includegraphics[width=0.6\linewidth]{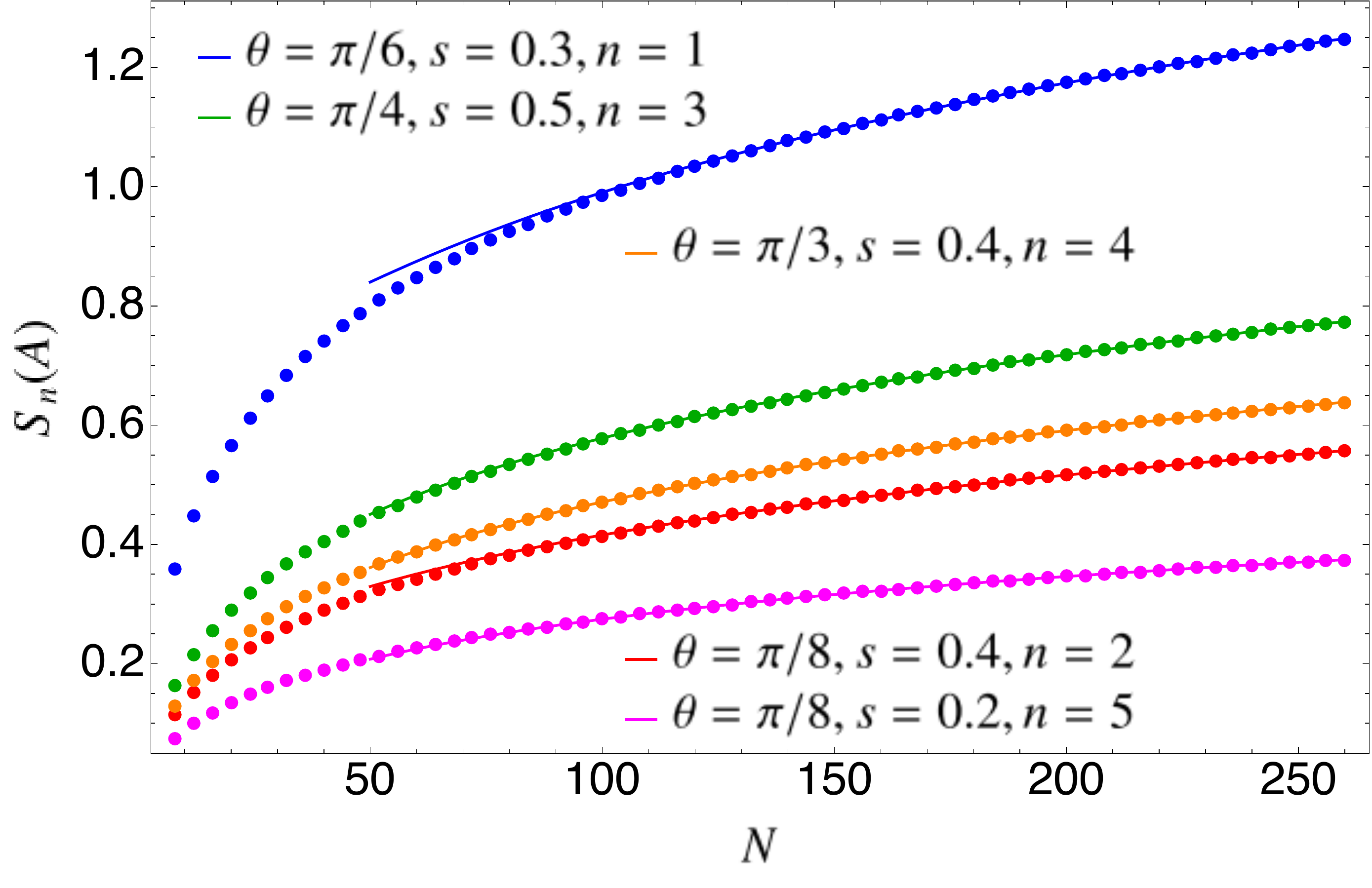}
    \caption{The R\'enyi entropies $S_n(A)$ for 4 bosonic species with $M_A=2$.
We choose different values of $s, \theta,n$ and we plot it as a function of the number of sites $N$. The lines show the curve $C_n (s, \theta) \log N + b_0 + b_1N^{-1/n}$ where the coefficients $b_i$ are fitted using the data for $N \geq 100$. The coefficients $C_n (s, \theta)$ are obtained by summing over the single-wire results, as explained in Eq. \eqref{eq:cal-final}}
    \label{fig:n0}
\end{figure}
Let us now consider the R\'enyi entropy between $M_A$ wires and the remaining $M-M_A$ ones. It is convenient to introduce the matrices $X_A$ and $P_A$ of  dimension $(M_AN,M_A N)$ obtained from $X,P$ in Eq. \eqref{eq:xil} by restricting $l,l'=1,\dots ,M_A$. In terms of these matrices, the R\'enyi entropy of $A$ is given by
\be
S_{n}(A) = -\frac{2}{1-n}\text{Tr} \log \l (\sqrt{X_A P_A}+1/2)^n - (\sqrt{X_A P_A}-1/2)^n \r,
\ee
where the prefactor $2$ comes from the fact that we have to consider the contribution of each real Harmonic chain. Recalling that $S_{AA}$ is the projected $S$-matrix in the subsystem $S$, it is relevant to observe that
\be
X_A P_A  = \frac{1}{4} + \frac{1-S^2_{AA}}{4}\otimes (X_{DD}-X_{ND})(P_{ND}-P_{DD}),
\ee
which is a straightforward consequence of $X_{ND}P_{ND} =X_{DD}P_{DD} =\frac{1}{4}$. One can show that this property implies 
\be
S_{n}(A) = \sum_{\mathcal{T}_a } S_{n,\mathcal{T}_a},
\ee
where $\mathcal{T}_a$ is an eigenvalue of $1-S^2_{AA}$. Therefore, the total R\'enyi entropy can be decomposed as a sum of single-wire contributions, in agreement with the CFT result in Eq. \eqref{eq:rcft}.
We can use the machinery developed in this section to test the validity of the CFT result for the logarithmic scaling of the R\'enyi entropy. 
We consider the case $M_A=M_B=2$ and the $S$ matrix is chosen of the form (we observe that the same choice has been done for the fermionic junctions in \cite{cmc-22})
\begin{equation}
S=U\begin{pmatrix}
-\sqrt{1-s^2} & -s & 0 & 0\\
-s & \sqrt{1-s^2} & 0 & 0\\
0 & 0 & -1 & 0 \\
0 & 0 & 0 & 1 \\
\end{pmatrix}U^{-1} , \quad U=\begin{pmatrix}
1 & 0 & 0 & 0 \\
0 & -\cos \theta &  -\cos \theta \sin \theta & \sin^2 \theta \\
0 & \sin \theta & -\cos^2 \theta & \cos \theta \sin \theta\\
0 & 0 & \sin \theta & \cos \theta
\end{pmatrix}.
\end{equation}
The numerical results are reported in Fig. \ref{fig:n0} finding a perfect agreement with the CFT result. \subsection{Entanglement negativity between two species}
\begin{figure}[t]
\centering
\subfigure
{\includegraphics[width=0.48\textwidth]{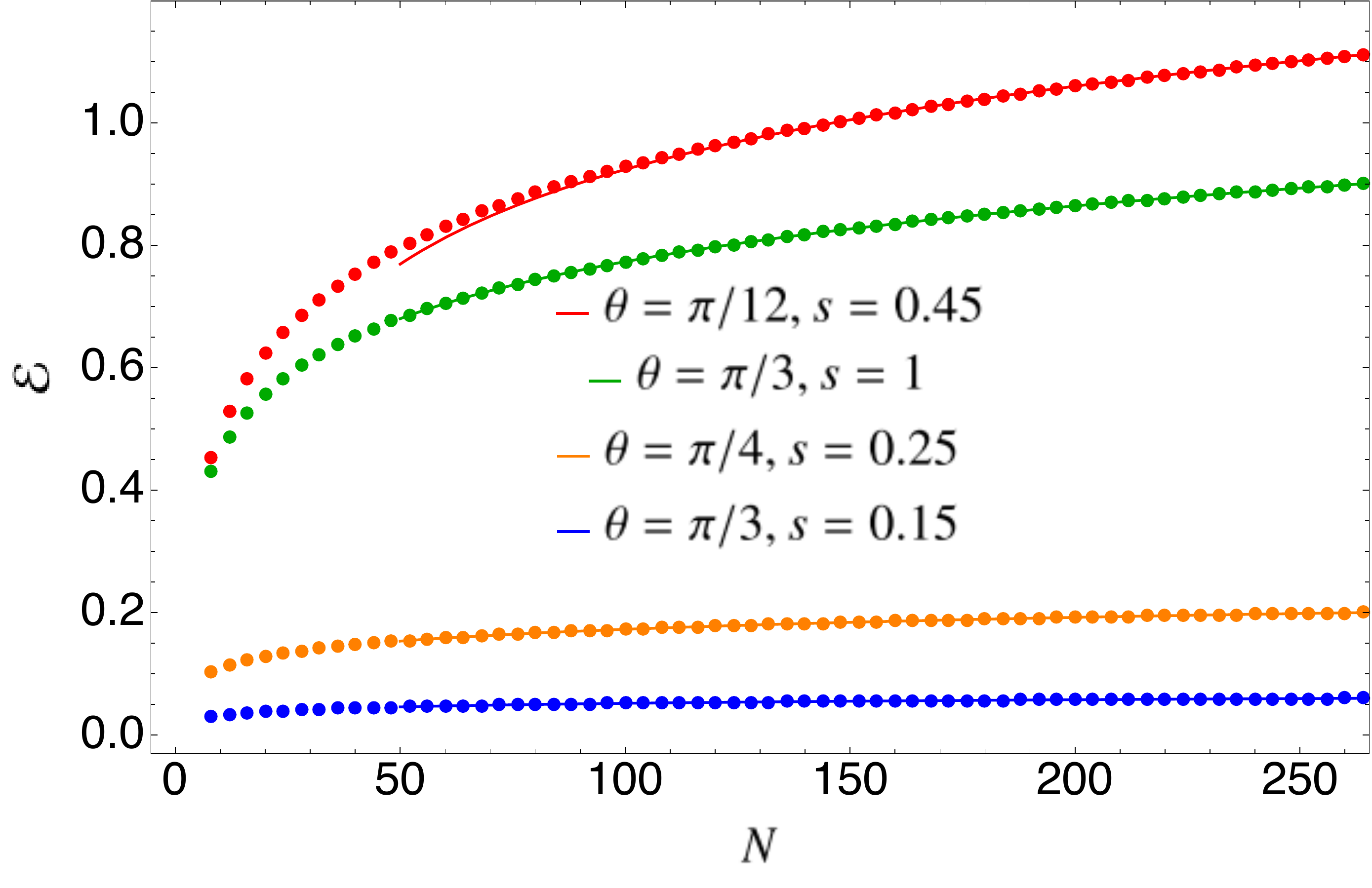}}
\subfigure
{\includegraphics[width=0.48\textwidth]{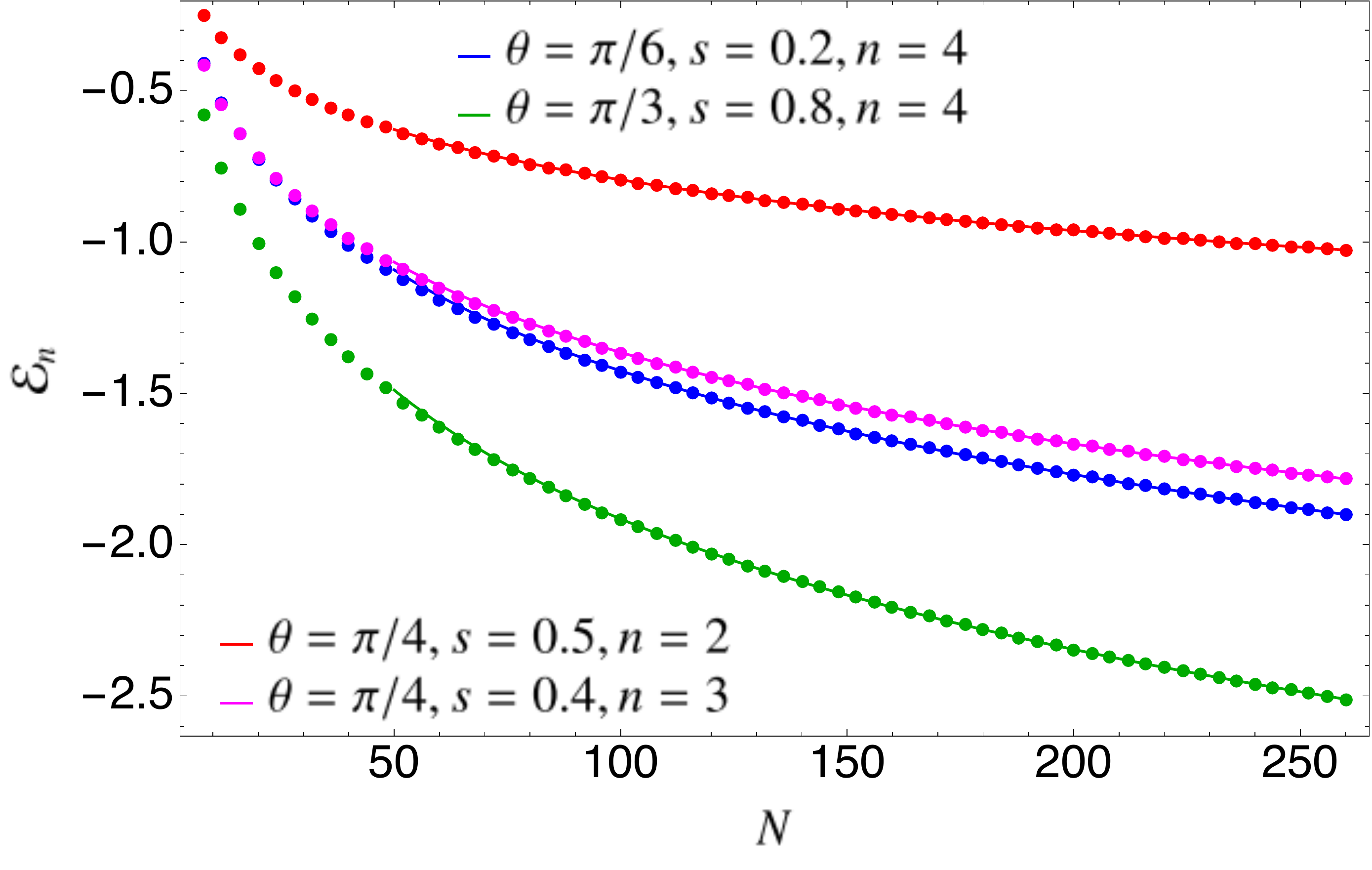}}
\caption{The R\'enyi negativities $\mathcal{E}_{n}$ in the three-wire junction for different values of $s,\theta$ and as a function of number of particles $N$. 
The lines show the curve $C_{n}(s,\theta)\log N+b_0+b_1N^{-1}$ where the coefficients $b_i$ are fitted using the data for $N \geq 80$. 
The prefactor $C_{n}(s,\theta)$ is given by Eq. \eqref{eq:negativity}.}\label{fig:n2}
\end{figure}
\begin{figure}[t]
\centering
	\includegraphics[width=0.5\linewidth]{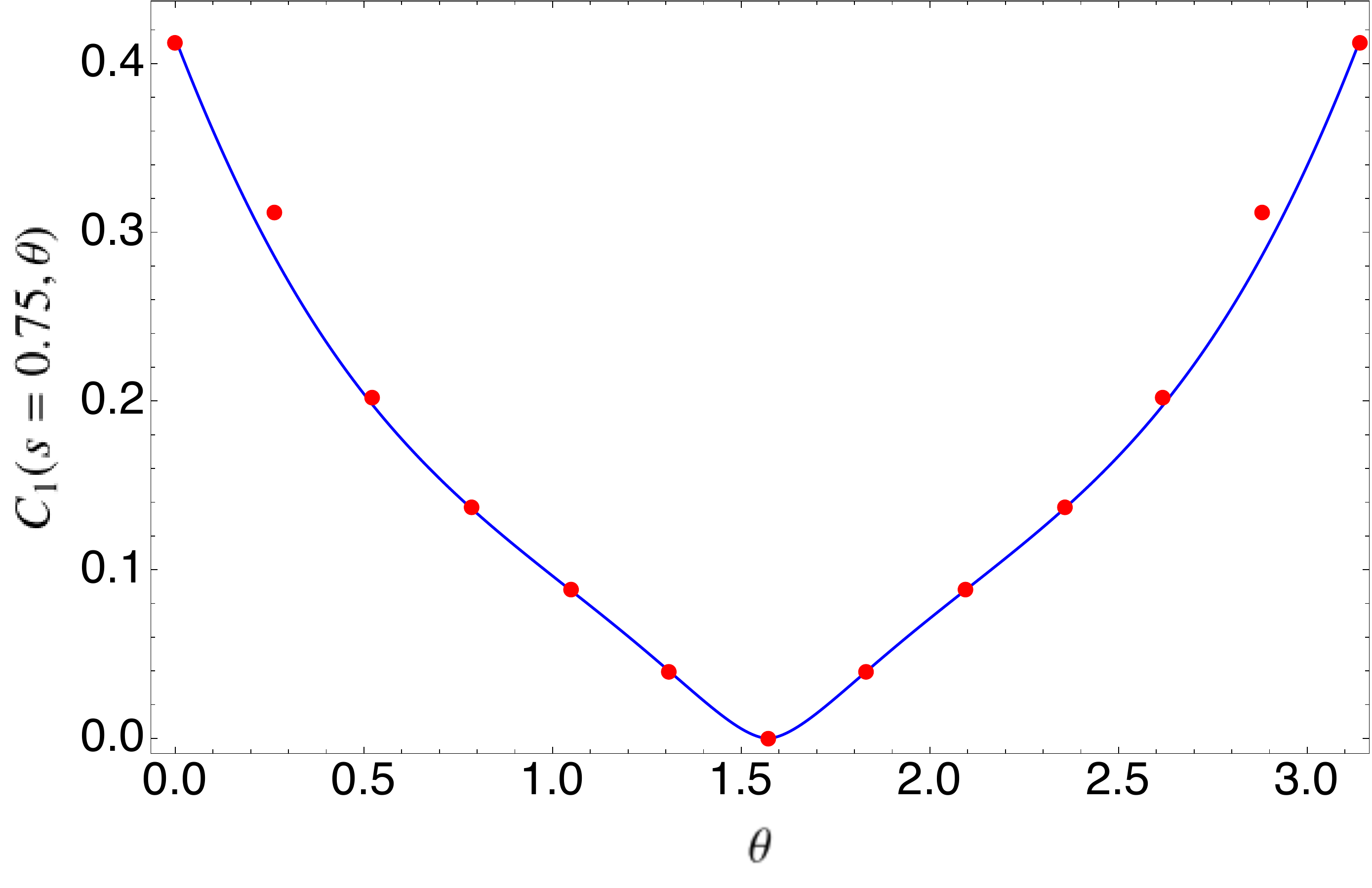}
    \caption{The coefficient of the logarithmic term of the negativity between two wires ($A$ and $B$) as a function of $\theta$, with fixed $s=0.75$. 
    The solid line corresponds to Eq. \eqref{eq:negativity} while the points have been obtained through a fit of the numerics with the form $a \log N+b_0+b_1 N^{-1}$. }
    \label{fig:n1}
\end{figure}

We now consider a tripartition $A\cup B \cup C$, where $A$ $(B)$ contains $M_A$ ($M_B$) wires, and we study
the entanglement negativity between $A$ and $B$.
To achieve this goal, we have to consider the
$(M_A+M_B)N,(M_A+M_B)N$ correlation matrices
$X_{A\cup B}, P_{A\cup B}$.
The partial transpose operation on $A$ has the net effect of changing the sign of the momenta corresponding
to the transposed subsystem. In other words, given the block decomposition
\be
P_{A \cup B} = \begin{pmatrix} P_{AA} & P_{AB}\\ P_{BA} & P_{BB}\end{pmatrix}
\ee
the partial transposition along $A$ produces
\be
\tilde{P}_{A \cup B} = \begin{pmatrix} P_{AA} & -P_{AB}\\ -P_{BA} & P_{BB}\end{pmatrix}.
\ee
In this way, one can express the $n$-th R\'enyi negativity between $A$ and $B$ as
\be\label{eq_NegDefect}
\mathcal{E}_{n}   = -2\text{Tr} \log\l \left( \sqrt{X_A \tilde{P}_A}+1/2\right)^n - \left(\sqrt{X_A \tilde{P}_A}-1/2\right)^n \r.
\ee
The replica limit is obtained by taking
\begin{equation}
  \mathcal{E}  = -2\text{Tr} \log\l \left| \sqrt{X_A \tilde{P}_A}+1/2\right| - \left|\sqrt{X_A \tilde{P}_A}-1/2\right| \r.  
\end{equation}
Eq. \eqref{eq_NegDefect} gives the R\'enyi negativities in terms of the correlation matrices that, once numerically evaluated, provides a test of the CFT 
results for the coefficient of the logarithmic growth  obtained in Section \ref{sec:negativity}. 
We test a three-wire junction with a two-parameter family of scattering matrices given by
\begin{equation}
S=U\begin{pmatrix}
-\sqrt{1-s^2} & -s & 0\\
-s & \sqrt{1-s^2} & 0 \\
0 & 0 & -1
\end{pmatrix}U^{-1} , \quad U=\begin{pmatrix}
1 & 0 & 0 \\
0 & -\cos \theta & \sin \theta \\
0 & \sin \theta & \cos \theta \\
\end{pmatrix}.
\end{equation}
Here $A$ and $B$ are the first two wires and we compute numerically the R\'enyi negativity for several values of $s,\theta, $ and $N$. In Fig. \ref{fig:n2} we benchmark the $N$ dependence of the R\'enyi negativity and, in particular, the prefactor of the logarithmic term in 
Eq. \eqref{eq:negativity} for different values of $n$ (rught panel) and for the replica limit $n_e \to 1$ (left panel).
In Fig. \ref{fig:n1} we test the coefficient of the logarithm obtained fixing $s=0.75$ and varying $\theta$;
for each pair $(s,\theta)$, we calculate numerically the negativity, for several values of $N$ up to 300. We use as fitting formula for the numerical results $a \log N+b_0+b_1 N^{-1}$. 
Fig. \ref{fig:n1} finally reports the best fit of $a$ as a function of $\theta$, finding good agreement with the analytic continuation done in Eq. \eqref{eq:negativity}. 
The minor discrepancy close to $\theta\sim0.2$ (and to the symmetric point) is due to finite size corrections that for small, but non-vanishing, $s$ are more relevant.

\section{Conclusions}\label{sec:conlusions}

In this manuscript we have investigated the entanglement properties of a one-dimensional system made by $M$ complex bosons coupled together at a single point. We consider two complementary subsets $A$ and $B$, of $M_A,M_B$ wires respectively, and we show that their R\'enyi entropy and negativity grow logarithmically in the system sizes. The universal predictions have been derived via a generalisation of a technique introduced in \cite{cmc-22}, relating the entanglement measures to $U(1)$-charged partition function in CFT.\\
We found that the entanglement among the wires depends only on the entries of a scattering matrix $S$, which describes the scattering among the distinct species. Importantly, our results for the free bosons differ explicitly from the ones derived for the free fermions in \cite{cmc-22}, showing that the two boundary theories cannot be simply related by bosonisation, a feature which was already observed in \cite{bddo-02}.
We mention the obvious fact that, for coupled wires of {\it real} massless scalars, both entanglement entropy and negativity are given by half of our results.\\
We discuss some possible outlooks. A possible generalisation would be the computation of the negativity in other interacting CFTs with a defect (e.g. WZW models, 3-state Potts model, $\mathbb{Z}_n$ parafermions), following, for example, the techniques of \cite{bbjs-16} valid for topological interfaces. Another interesting direction regards the dynamics of these  systems with defects in CFT after a global/local quantum quench (see, e.g., Refs. \cite{cc-16,wwr-18,ddb-21,rfgc-22}). A last direction could be the investigation of  the entanglement of other bipartitions, like a subsystem made of an interval of length $\ell$ belonging to a single wire (of length $L$) and attached to the boundary, exploiting for example the methods of Ref. \cite{tm-21}.

\section*{Acknowledgements}
All authors acknowledge support from ERC under Consolidator grant number 771536 (NEMO).

\begin{appendix}

\section{Useful identities}
\subsection{Expectation value of Gaussian operators}

Here we prove a useful identity valid for the vacuum expectation value of Gaussian operators
\be
\bra{0} \exp\l \mathcal{O}'_{jj'} \Phi^{j}_{k} \bar{\Phi}^{j'}_{k}\r\exp\l \mathcal{O}_{jj'} \bar{\Phi}^j_{-k}\Phi^{j'}_{-k}\r\ket{0} = \text{det}\l 1-\mathcal{O}'\mathcal{O} \r^{-1}.
\label{eq_BoundOverlap}
\ee
We denoted by $\Phi^j_{\mp k}$ the creation/annihilation operators associated to the $k$-th left mode of the $j$-th bosonic specie, and similarly we use the symbol $\Phi^j_{\mp k}$ for the right modes. We consider $\mathcal{O}$ and $\mathcal{O}'$ as two generic $M\times M$ matrices, keeping implicit the sum over the indices $j,j'$ in Eq. \eqref{eq_BoundOverlap}. We assume further that $k\neq 0$, so that the zero-mode of the boson does not appear here, and we use the following convention for the normalisation of the modes
\be
[\Phi^{j}_{k},\Phi^{j'}_{-k'}] = [\bar{\Phi}^{j}_{k},\bar{\Phi}^{j'}_{-k'}] = \delta_{jj'} \delta_{kk'}, \quad k,k'>0.
\ee
It is possible to highlight a simple case where \eqref{eq_BoundOverlap} holds. For instance, we take a single bosonic specie ($M=1$), we suppress the index $j$, and then $\mathcal{O}'$ and $\mathcal{O}$ are just numbers. Expanding the exponential as a power series in \eqref{eq_BoundOverlap} we get
\be
\bra{0} \exp\l \mathcal{O}' \Phi_{k} \bar{\Phi}_{k}\r\exp\l \mathcal{O} \bar{\Phi}_{-k}\Phi_{-k}\r\ket{0} = \bra{0}\sum^{\infty}_{n=0}\sum^{\infty}_{m=0} \frac{(\mathcal{O})^n(\mathcal{O}')^m}{n! m!} (\Phi_{k})^n (\bar{\Phi}_{k})^n (\bar{\Phi}_{-k})^m(\Phi_{-k})^m\ket{0}.
\label{eq_BoundOverlap1}
\ee
Via Wick theorem, one easily shows that
\be
\bra{0}(\Phi_{k})^n(\Phi_{-k})^m\ket{0} = \bra{0}(\bar{\Phi}_{k})^n(\bar{\Phi}_{-k})^m\ket{0} = m! \delta_{m,n}.
\ee
Inserting this relation back into \eqref{eq_BoundOverlap1} one gets
\be
\bra{0} \exp\l \mathcal{O}' \Phi_{k} \bar{\Phi}_{k}\r\exp\l \mathcal{O} \bar{\Phi}_{-k}\Phi_{-k}\r\ket{0} = \sum^{\infty}_{n=0}(\mathcal{O}'\mathcal{O})^n = \frac{1}{1-\mathcal{O}'\mathcal{O}},
\ee
which is compatible with the general case in Eq. \eqref{eq_BoundOverlap}. Clearly for $M\geq 1$, whenever the matrices $\mathcal{O}$ and ${\mathcal{O}'}$ commute, one can diagonalise them simultaneously and apply the previous argument to show the validity of \eqref{eq_BoundOverlap}.\\
A general proof of \eqref{eq_BoundOverlap} is given through Gaussian integrals over complex variables. We start from the basic property \cite{zj-89}
\be
\int \frac{dz d\bar{z}}{\pi} \exp(-z\bar{z}) =1,
\ee
which is the building block of the forecoming relations. For each $j$ we introduce a complex variable $z_j$, and we express the operator $\exp\l \mathcal{O}_{jj'} \bar{\Phi}^j_{-k}\Phi^{j'}_{-k}\r$ as follows
\be
\exp\l \mathcal{O}_{jj'} \bar{\Phi}^j_{-k}\Phi^{j'}_{-k}\r = \int dz d\bar{z}\exp\l -z_j\bar{z}_j +z_j \mathcal{O}_{jj'}\Phi^{j'}_{-k} +\bar{\Phi}^{j}_{-k}\bar{z}_j \r,
\label{eq:Gauss_rep1}
\ee
with $dzd\bar{z}$ being a short-hand notation for the normalised measure
\be
dzd\bar{z} \equiv \frac{dz_1d\bar{z}_1\dots dz_Md\bar{z}_M}{\pi^M},
\ee
and the sum over $j,j'$ is implicit. Similarly, we represent
\be
\exp\l \mathcal{O}'_{jj'} \Phi^{j}_{k} \bar{\Phi}^{j'}_{k}\r = \int dw d\bar{w}\exp\l -\bar{w}_jw_j + \Phi^j_k \mathcal{O}'_{jj'}w_{j'}+\bar{w}_j\bar{\Phi}^j_{k}\r,
\label{eq:Gauss_rep2}
\ee
after the introduction of a set of complex variables $\{w_j\}_{j=1,\dots,M}$. We now use \eqref{eq:Gauss_rep1} and \eqref{eq:Gauss_rep2} and we write
\be
\begin{split}
\bra{0} \exp\l \mathcal{O}'_{jj'} \Phi^{j}_{k} \bar{\Phi}^{j'}_{k}\r\exp\l \mathcal{O}_{jj'} \bar{\Phi}^j_{-k}\Phi^{j'}_{-k}\r\ket{0} = \int dz d\bar{z} dw d\bar{w} \exp\l
-z_j\bar{z}_j-\bar{w}_jw_j
\r\\
\bra{0}\exp\l \bar{w}_j\bar{\Phi}^j_{k} \r \exp\l \bar{\Phi}^{j}_{-k}\bar{z}_j\r\ket{0}\bra{0}\exp\l \Phi^j_k \mathcal{O}'_{jj'}w_{j'}\r\exp\l z_j \mathcal{O}_{jj'}\Phi^{j'}_{-k} \r\ket{0}=\\
\int dz d\bar{z} dw d\bar{w}\exp\l
-z_j\bar{z}_j-\bar{w}_jw_j+\bar{w}_j\bar{z}_j +z_j (\mathcal{O}\mathcal{O}')_{jj'}w_{j'} \r,
\end{split}
\label{eq:Bos_Integ}
\ee
where the basic properties of bosonic coherent states have been used (see \cite{zj-89}). As a final step, we perform the evaluation of the integral through the introduction of a $4M$-dimensional vector $\Theta$ of complex variables as follows
\be
\Theta = \begin{pmatrix}z \\ \bar{w} \\ \bar{z} \\ w \end{pmatrix}.
\ee
In this way, we express \eqref{eq:Bos_Integ} concisely as
\be
\bra{0} \exp\l \mathcal{O}'_{jj'} \Phi^{j}_{k} \bar{\Phi}^{j'}_{k}\r\exp\l \mathcal{O}_{jj'} \bar{\Phi}^j_{-k}\Phi^{j'}_{-k}\r\ket{0} = \int d\Theta \exp\l -\frac{1}{2}\Theta^T \tilde{\mathcal{O}}\Theta\r,
\ee
where $\tilde{\mathcal{O}}$ is the $4M\times 4M$ symmetric matrix given by
\be
\tilde{\mathcal{O}} = \begin{pmatrix} 0 & 0 & 1 & -\mathcal{O}\mathcal{O}' \\ 0 & 0 & -1 & 1 \\ 1 & -1 & 0 & 0\\ -(\mathcal{O}\mathcal{O}')^T & 1 & 0 & 0\end{pmatrix}.
\ee
One can perform the Gaussian integral over $\Theta$, which gives $\text{det}\l \tilde{\mathcal{O}}\r^{-1/2}$, and we express the final result as follows 
\be
 \text{det}^{-1/2}\l \tilde{\mathcal{O}} \r = \text{det}^{-1}\begin{pmatrix}  1 & -\mathcal{O}\mathcal{O}' \\  -1 & 1 \end{pmatrix} = \text{det}^{-1}(1-\mathcal{O}'\mathcal{O}).
\ee
In this way, we complete the proof of Eq. \eqref{eq_BoundOverlap} in the most general case, which is the main result of this section.

\subsection{Jacobi Theta functions and Dilogarithm }\label{sec_Jacobi}

In this section we review the definition and some basic properties of the Jacobi Theta functions and of the dilogarithm. We express the Jacobi Theta function $\theta_1(z,q)$ as the following infinite product \cite{gr-94}
\be
\theta_1(z,q) = 2\sin(\pi z)q^{1/4} \prod_{m=1}^\infty 
(1-q^{2m})\left( 1 - 2 \cos(2 \pi z)q^{2m}+q^{4m}\right).
\label{eq:Theta1_def}
\ee
We want to investigate the singular limit $q\rightarrow 1^-$ of \eqref{eq:Theta1_def}, focusing on its explicit $z$-dependence. To do
so, we approximate the infinite product as an integral 
\begin{multline}
\log \frac{\theta_1(z,q)}{\sin(\pi z)} \simeq \int^{\infty}_{0}dx \ [\log(1-e^{i2\pi z}q^{2x})+\log(1-e^{-i2\pi z}q^{2x})] + \text{const.} =\\
-\frac{1}{2\log q}\int^{1}_{0}\frac{dt}{t}[\log(1-e^{i2\pi z}t)+\log(1-e^{-i2\pi z}t)] + \text{const.}=\\
\frac{1}{2\log q}\l \text{Li}_2(e^{i2\pi z})+\text{Li}_2(e^{-i2\pi z})\r + \text{const.},
\end{multline}
where the irrelevant additive constant does not depend on $z$. In the derivation of the previous formula we make use of the integral representation of the dilogarithm
\be
\text{Li}_2(z) = -\int^{1}_0\frac{dt}{t}\log(1-zt).
\ee
From the definition in \eqref{eq:Theta1_def} it is clear that $\theta_1(z,q)/\sin(\pi z)$ is periodic under
\be
z\rightarrow z+1,
\ee
and for this reason we focus only on the region of parameter $z \in [0,1]$. In this case one can show that
\be
\log \frac{\theta_1(z,q)}{\sin(\pi z)} \simeq -\frac{\pi^2(z-z^2)}{\log q}+\text{const.}, \quad z \in [0,1],
\ee
up to an irrelevant $z$-independent additive constant.\\
As a final remark, we emphasise that $\theta_1(z,q)$ goes linearly to zero as $z\rightarrow 0$. This implies that the following limit
\be
\underset{z\rightarrow 0}{\lim} \ \frac{\theta_1(z,q)}{\sin(\pi z)}
\ee
is nonsingular, and its value is just a (non-zero) number.

\subsection{Trigonometric identities}\label{sec:trig}

Here we derive some algebraic formula which are useful for the evaluation of the analytical continuations of the entanglement measures. The first identity we consider is the following \cite{gr-94}
\be
\prod^{n-1}_{p=0}(x-e^{i2\pi p/n}y) = x^n-y^n, \quad n \in \mathbb{N},
\ee
which comes from the factorisation of the polynomial $x^n-y^n$. A straightforward consequence of the previous relation is
\be
\prod^{n-1}_{p=0}(x^2-2\cos \frac{2\pi p}{n}xy + y^2) = \prod^{n-1}_{p=0}(x-e^{i2\pi p/n}y)\prod^{n-1}_{p=0}(x-e^{-i2\pi p/n}y) = (x^n-y^n)^2.
\label{eq:TrigId0}
\ee
From the previous formula, we will derive
\be
\prod^{n_e-1}_{p=0}\l x- \cos \frac{2\pi p}{n_e} y\r = \l \l\frac{x+\sqrt{x^2-y^2}}{2}\r^{n_e/2} - \l\frac{x-\sqrt{x^2-y^2}}{2}\r^{n_e/2}\r^2,
\label{eq:TrigId}
\ee
valid for every even integer $n_e$. To proceed with the proof of Eq. \eqref{eq:TrigId}, we firstly write
\be
\prod^{n_e-1}_{p=0}\l x- \cos \frac{2\pi p}{n_e} y\r   =  \prod^{n_e-1}_{p=0}\l x_1^2-2\cos \frac{2\pi p}{n_e}x_1y_1 + y_1^2\r ,
\ee
where $x_1,y_1$ are defined by
\be
\begin{cases} x_1^2+y_1^2 = x, \\ 2x_1y_1 = y\end{cases}.
\ee
The only solution to this system of equations is
\be
\begin{cases}x_1^2 = \frac{x+\sqrt{x^2-y^2}}{2},\\ y_1^2 = \frac{x-\sqrt{x^2-y^2}}{2},  \end{cases}
\ee
up to the exchange $x_1 \leftrightarrow y_1$. Since $n_e$ is even, it holds
\be
x_1^{n_e} = \l x^2_1 \r^{n_e/2}, \quad y_1^{n_e} = \l y^2_1 \r^{n_e/2}.
\ee
Replacing $x_1,y_1$ as a function of $x,y$ and making use of Eq. \eqref{eq:TrigId0}, one finally reaches to \eqref{eq:TrigId}.\\
An immediate application of the formula we derived so far is the evaluation of the product
\be
\prod^{n_e-1}_{p=0}\l x+ \cos \frac{2\pi p}{n_e} y + \cos^2 \frac{2\pi p}{n_e} z\r.
\ee
Indeed, through a factorisation of the term inside the parenthesis as second order polynomial in $\cos \frac{2\pi p}{n_e}$, one can split the product in two products and each of them can be computed from Eq. \eqref{eq:TrigId}.

\end{appendix}

\end{document}